\documentstyle[12pt]{article}

\begin{document}

\title{\bf The relativistic causal Newton gravity law \hskip 2cm vs. general relativity}

\author{Yury M. Zinoviev\thanks{Electronic mail: zinoviev@mi.ras.ru}}

\date{}
\maketitle

{\it Steklov Mathematical Institute, Gubkin Street 8, 119991, Moscow,
Russia}

\vskip 0.5cm

\noindent The equations of the relativistic causal Newton gravity law for
the planets of the solar system are studied in the approximation when the
Sun rests at the coordinates origin and the planets do not interact between
each other. The planet orbits of general relativity are also studied in
the same approximation.

\vskip 0.5cm

\section{I. INTRODUCTION}
\setcounter{equation}{0}

Poincar\'e${}^{1}$ tried to find a modification of Newton gravity law:
"In the paper cited Lorentz${}^{2}$ found it necessary to supplement his
hypothesis in such a way that the relativity postulate could be valid for
other forces in addition to the electromagnetic ones. According to his idea,
because of Lorentz transformation (and therefore because of translational
movement) all forces behave like electromagnetic (despite their origin).

"It turned out to be necessary to consider this hypothesis more attentively
and to study the changes it makes in the gravity laws in particular. First,
it obviously enables us to suppose that the gravity forces propagate not
instantly, but at the speed of light. One could think that this is a sufficient
for rejecting such a hypothesis, because Laplace has shown that this cannot occur.
But, in fact, the effect of this propagation is largely balanced by some other
circumstance, hence, there is no any contradiction between the law proposed and
the astronomical observations.

"Is it possible to find a law satisfying the condition stated by Lorentz and
at the same time reducing to Newton law in all the cases where the velocities
of the celestial bodies are small to neglect their squares (and also the products
of the accelerations and the distance) compared with the square of the speed of
light?"

The relativistic Newton gravity law was proposed in Ref. 3
\begin{equation}
\label{1.1} \frac{d}{dt} \left( \left( 1 - \frac{1}{c^{2}} \Biggl| \frac{d{\bf
x}_{k}}{dt} \Biggr|^{2} \right)^{- 1/2} \frac{dx_{k}^{\mu}}{dt}
\right) = - \eta^{\mu \mu} \sum_{\nu \, =\, 0}^{3} \frac{1}{c}
\frac{dx_{k}^{\nu}}{dt} \sum_{j\, =\, 1,2,\, j \neq k} F_{j;\, \mu
\nu}(x_{k},x_{j}),
\end{equation}
$k = 1,2$, $\mu = 0,...,3$. The world line $x_{k}^{\mu}(t)$ satisfies
the condition $x_{k}^{0}(t) = ct$; $c$ is the speed of light; the diagonal
$4\times 4$ - matrix $\eta^{\mu \nu} = \eta_{\mu \nu}$, $\eta^{00} = -
\eta^{11} = - \eta^{22} = - \eta^{33} = 1$; the strength
$F_{j;\, \mu \nu}(x_{k},x_{j})$ is expressed through the vector potential
\begin{equation}
\label{1.2} F_{j;\, \mu \nu}(x_{k},x_{j}) = \frac{\partial
A_{j;\, \nu}(x_{k},x_{j})}{\partial x_{k}^{\mu}} - \frac{\partial
A_{j;\, \mu}(x_{k},x_{j})}{\partial x_{k}^{\nu}},
\end{equation}
$$
A_{j;\, \mu}(x_{k},x_{j}) = 4\pi m_{j}G \eta_{\mu \mu} \int dt
e_{0}(x_{k} - x_{j}(t)) \frac{dx_{j}^{\mu}(t)}{dt} =
$$
\begin{equation}
\label{1.3} \eta_{\mu \mu} m_{j}G\left(
\frac{d}{dt^{\prime}} x_{j}^{\mu}(t^{\prime})\right) \left( c|{\bf
x}_{k} - {\bf x}_{j}(t^{\prime})| - \sum_{i\, =\, 1}^{3} (x_{k}^{i}
- x_{j}^{i}(t^{\prime})) \frac{d}{dt^{\prime}}
x_{j}^{i}(t^{\prime})\right)^{- 1},
\end{equation}
$$
t^{\prime} = c^{- 1}(x_{k}^{0} - |{\bf x}_{k} - {\bf
x}_{j}(t^{\prime})|), \, \,
e_{0}(x) = -\, (2\pi)^{- 1} \theta (x^{0})\delta ((x,x));
$$
the gravitation constant $G = (6.673 \pm 0.003)\cdot 10^{-
11}m^{3}kg^{- 1}s^{- 2}$ and $m_{j}$ is the $j$ body mass. The
distribution $e_{0} (x) \in S^{\prime} ({\bf R}^{4})$ with support in the
closed upper light cone is the fundamental solution of the wave equation.
It is unique. The distribution $e_{0}(x)$ is not a regular function. The
relativistic Newton law is based on the fundamental solution $e_{0}(x)$ of
the wave equation similar as Newton law is based on the fundamental solution
$- (4\pi)^{- 1} |{\bf x}|^{- 1}$ of Laplace equation. The vector potential
(\ref{1.3}) was proposed by Li\'enard (1898) and Wiechert (1900). The vector
potential (\ref{1.3}) is the relativistic version of Coulomb potential.

Newton gravity law requires the instant propagation of the force action. The special
relativity requires that the propagation speed does not exceed the speed of light. If the
propagation speed is independent of gravitating body speed, then it is equal to that of
light. The vector potential (\ref{1.3}) depends not on its simultaneous positions and
speeds but on the positions and the speeds at the time moments $t$ and $t^{\prime}$ which
differ from each other in the time interval $c^{- 1}|{\bf x}_{k}(t) - {\bf x}_{j}(t^{\prime})|$
needed for light covering the distance between the physical points ${\bf x}_{k}(t)$ and
${\bf x}_{j}(t^{\prime})$. The equations (\ref{1.1}) - (\ref{1.3}) satisfy the causality
condition: some event in the system can influence the evolution of the system in the
future only and can not influence the behavior of the system in the past, in the time
preceding the given event.

The equations (\ref{1.1}) - (\ref{1.3}) are the relativistic causal version of Newton
gravity law equations. Sommerfeld (Ref. 4, Sec. 38):
"The question may arise: what is the relativistic form of Newton gravity law? If the law
is supposed to have a vector form, this question is wrong. The gravitational field is not a
vector field. It has the incomparably complicated tensor structure." It seems the reason why
the relativistic Newton gravity law${}^{1}$ was not studied.

For the resting body world line $x_{j}^{0} (t) = ct$, ${\bf x}_{j} (t) = \hbox{const}$
the vector potential (\ref{1.3}) is
\begin{equation}
\label{1.4} A_{j;\, 0}(x_{k}, x_{j}) = m_{j}G|{\bf x}_{k} - {\bf
x}_{j}(c^{- 1}x_{k}^{0})|^{- 1},\, \, A_{j;\, i}(x_{k},x_{j}) = 0, \,
\, i = 1,2,3.
\end{equation}
$99.87\%$ of the solar system total mass belongs to the Sun. We consider the Sun resting
at the coordinates origin. The substitution of the vector potential (\ref{1.4}) for the Sun 
world line $x_{j}^{0} (t) = ct$, $x_{j}^{i} (t) = 0$, $i = 1,2,3$ into the equation (\ref{1.1}) 
yields
\begin{equation}
\label{1.13} \frac{d}{dt} \left( \left( 1 - \frac{1}{c^{2}} \Biggl| \frac{d{\bf
x}}{dt}\Biggr|^{2}\right)^{- 1/2} \frac{dx^{i}}{dt} \right) =
- \frac{m_{10}Gx^{i}}{|{\bf x}|^{3}},\, \, i = 1,2,3.
\end{equation}
$m_{10}$ is the Sun mass. The right-hand side of the equation (\ref{1.13}) coincides with
the right-hand side of the Newton gravity law equation for a planet. We neglect the 
interaction between the planets. The equation (\ref{1.13}) is solved in Ref. 3. We choose 
the third axis to be orthogonal to the orbit plain
\begin{equation}
\label{1.11}
x^{1}(t) = r(t) \cos \phi (t),\, \, x^{2}(t) = r(t) \sin \phi (t),\, \,
x^{3}(t) = 0.
\end{equation}
The orbit radius $r(t)$ is given by
\begin{equation}
\label{1.5}
\frac{a(1 - e^{2})}{r(t)} = 1 + e \cos
\gamma (\phi (t) - \phi_{0} ).
\end{equation}
$\phi (t)$ is the orbit angle, $\phi_{0}$ is the perihelion orbit angle, $e$ is the
planet orbit eccentricity and $a$ is the "ellipse" (\ref{1.5}) major "semi - axis".
In the Section II the time dependence of the orbit radius $r(t)$ is determined. We
get the precession coefficient
\begin{equation}
\label{1.6}
\gamma \approx 1 - \frac{\omega^{2} a^{2}}{2(1 - e^{2})c^{2}}.
\end{equation}
$\omega = 2\pi T^{- 1}$ is the mean "angular frequency" and $T$ is the planet "period".
According to (Ref. 5, Chap. 25, Sec. 25.1, Appendix 25.1) for Mercury
$\omega^{2} a^{3}c^{- 2} = 1477m$, $a = 0.5791\cdot 10^{11}m$, $e = 0.21$ and
$2^{- 1}\omega^{2}a^{2} c^{- 2}(1 - e^{2})^{- 1} \approx 1.3341 \cdot 10^{- 8}$.
The advance of Mercury's perihelion, observed from the Sun, is
$(\gamma^{- 1} - 1) \cdot 360^{o}$ per "period" of Mercury. The advance of Mercury's
perihelion, observed from the Sun, is  $(\gamma^{- 1} - 1) \cdot 360\cdot 415
\cdot 3600^{''} \approx 2^{- 1}\omega^{2}a^{2} c^{- 2}(1 - e^{2})^{- 1} \cdot
360\cdot 415 \cdot 3600^{''} \approx 7^{''}.175$ per century ($415$ "periods" of Mercury).
$1^{o} = 60^{'} = 3600^{''}$. The advance of Mercury's perihelion, observed by the
astronomers from the Earth, is $5599^{''}.74 \pm 0^{''}.41$ per century (Ref. 5, Chap. 40,
Sec. 40.5, Appendix 40.3). By using Newton gravity law it is possible to calculate the advance
of Mercury's perihelion caused by the non-inertial system connected with the Earth.
It turns out to be $5025^{''}.645 \pm 0^{''}.50$ per century (Ref. 5, Chap. 40, Sec. 40.5,
Appendix 40.5). By using Newton gravity law it is possible to calculate the advance of
Mercury's perihelion caused by the gravity of other planets. It turns out to be
$531^{''}.54 \pm 0^{''}.68$ per century (Ref. 5, Chap. 40, Sec. 40.5, Appendix 40.5). The
rest advance of Mercury's perihelion $5599^{''}.74 - 5025^{''}.645 - 531^{''}.54 \approx 42^{''}.56$
per century can not be explained by the disturbing forces. It is not obvious that we can
add the advance of Mercury's perihelion obtained for the orbits in Newton gravity theory
and the advance of Mercury's perihelion $7^{''}.175$, observed from the Sun and obtained
for the orbits (\ref{1.11}) - (\ref{1.6}). In our opinion for the experimental verification
of the relativistic causal Newton gravity law (\ref{1.1}) - (\ref{1.3}) it is necessary to
obtain the advance of Mercury's perihelion $5599^{''}.74$, observed from the Earth, by
making use of the relativistic causal Newton gravity law (\ref{1.1}) - (\ref{1.3}) without
Newton gravity theory.

In the Section II we study the orbits (\ref{1.11}) - (\ref{1.6}) of Mercury and of the
Earth and show that the value of the Mercury's perihelion advance, observed from the
Earth, depends on the perihelion angle $\phi_{0}$ of the Mercury orbit and on the
perihelion angle $\phi_{0}$ of the Earth orbit.

Kepler (Astronomia nova seu physica coelestis, tradita commentariis de motibus 
stellae Martis ex obsevationibus Tychonis Brahe. MDCIX) found that the planet orbits are 
elliptic in the coordinate system where the Sun rests. Kepler used Tycho Brahe's 
astronomical observations (1580-1597). Due to Brahe, the 
Mars orbit deviation from the circular orbit was $8'$. Ptolemaeus and Copernicus had the
instrument precision $10'$. Brahe had the instrument precision $2'$. The
intensive astronomic observations from the middle of the XIX century and the
radio-location after 1966 discovered the advances of orbit perihelion for
different planets. Is the orbit (\ref{1.11}) - (\ref{1.6}) consistent with the
observable Mercury's orbit?  Clemence${}^{6}$:
"Observations of Mercury are among the most difficult in positional astronomy. They
have to be made in the daytime, near noon, under unfavorable conditions of the
atmosphere; and they are subject to large systematic and accidental errors arising
both from this cause and from the shape of the visible disk of the planet. The
planet's path in Newtonian space is not an ellipse but an exceedingly complicated
space-curve due to the disturbing effects of all of the other planets. The
calculation of this curve is a difficult and laborious task, and significantly
different results have been obtained by different computers."

By making use of Hamilton-Jacobi equation Boguslavskii (Ref. 7, P. 233 - 403. Boguslavskii
used the German transcription: Boguslawski.) solved the equation (\ref{1.13}) and obtained the
orbit formula (\ref{1.5}). Boguslavskii did not calculate all integrals needed for Hamilton-Jacobi
equation and did not obtain the time dependence of the orbit (\ref{1.11}), (\ref{1.5}) radius.
Boguslavskii (Ref. 7, P. 386): "Since any material point mass changes in the special relativity
according to the same law as the electron mass does, Einstein tried to apply the theory described 
above for the explanation of the part of Mercury's movement which can not be explained by the 
disturbing forces. However, the movement calculated due to the formula (\ref{1.5}) turned out to 
be six times less than the observable movement. Einstein${}^{8}$ obtained the correct explanation 
by means of his general relativity principle containing the new gravity theory." By making use
of the formulas (\ref{1.5}), (\ref{1.6}) Einstein obtained probably the estimate of advance
of Mercury's perihelion, observed from the Sun, $42^{''}.56:6 \approx 7^{''}.09$ per century.
Could Boguslavskii have learned Einstein's calculation of Mercury's movement by means of the 
formulas (\ref{1.5}), (\ref{1.6})? In 1913 - 1914 he worked in G\"ottingen University 
together with Max 
Born (Ref. 7, P. 9 - 17). 

The metric (Ref. 9, Chap. 38, relation (38.8))
\begin{equation}
\label{4.17}
(ds)^{2} = \left( 1 - 2 \frac{m_{10}G}{rc^{2}} \right) c^{2}(dt)^{2} -
\left( 1 - 2 \frac{m_{10}G}{rc^{2}} \right)^{- 1} (dr)^{2} - r^{2}(d\theta)^{2} -
r^{2}\sin^{2} \theta (d\phi)^{2},
\end{equation}
$$
r = ((x^{1})^{2} + (x^{2})^{2} + (x^{3})^{2})^{1/2}, \, \,
\theta = \arctan [x^{3}/((x^{1})^{2} + (x^{2})^{2})^{1/2}], \, \,
\phi = \arctan (x^{2}/x^{1}),
$$
is a solution of Einstein's equations: $R_{\mu \nu} (x) - (1/2)g_{\mu \nu} (x)R(x) = 0$,
$x \neq 0$. The solution (\ref{4.17}) was obtained by Schwarzschild (1916). Eddington${}^{9}$
considers the Sun resting at the coordinates origin and neglects the interaction between the 
planets. The Sun gravitational field is described by the metric (\ref{4.17}).
A planet orbit is a solution of the geodesic equation (Ref. 9, Chap. 39, equation (39.1)) for
the metric (\ref{4.17}). The constant angle $\theta = \pi /2$ satisfies the geodesic equation
(39.2) from Ref. 9, Chap. 39. Choose the coordinates in such a way that a planet moves in the
plane $\theta = \pi /2$. The geodesic equations (39.1) from Ref. 9, Chap. 39 for the metric
(\ref{4.17}) imply the following equations (Ref. 9, Chap. 39, equations (39.61), (39.62))
\begin{equation}
\label{4.18} \frac{d^{2}}{d\phi^{2}} \, \, \frac{1}{r} + \frac{1}{r} - \frac{m_{10}G}{c^{2}h^{2}}
- \frac{3m_{10}G}{c^{2}r^{2}} = 0, \, \, r^{2}\frac{d\phi}{ds} = h = \hbox{const}.
\end{equation}

For Venus the orbit eccentricity $e = 0.007$. Venus moves along the approximately circular orbit
with approximately constant radius $a$ and with approximately constant angular frequency $\omega$.
The substitution of the relation (\ref{1.11}) for $r(t) = a$, $\phi (t) = \omega t - \phi_{0}$
into the Newton gravity law for Venus and the Sun yields the third Kepler law
\begin{equation}
\label{1.10}
m_{10}G = \omega^{2} a^{3}.
\end{equation}

According to (Ref. 5, Chap. 25, Sec. 25.1, Appendix 25.1) for Mercury, Venus, the Earth,
Mars and Saturn  $\omega^{2} a^{3}c^{- 2} = 1477m$, for Jupiter and Neptune
$\omega^{2} a^{3}c^{- 2} = 1478m$, for Uranus $\omega^{2} a^{3}c^{- 2} = 1476m$, for
Pluto $\omega^{2} a^{3}c^{- 2} = 1469m$. The value
$m_{10}Ga^{- 1}c^{- 2} = \omega^{2} a^{2}c^{- 2}$ is negligeable for any planet. For
the nearest planet to the Sun, Mercury  $a = 0.5791\cdot 10^{11}m$. We substitute the
function (\ref{1.5}) into the first equation (\ref{4.18}) and multiply the obtained
equality by $a(1 - e^{2})$
\begin{equation}
\label{4.30} 1 - \frac{m_{10}Ga(1 - e^{2})}{h^{2}c^{2}}
+  \left( 1 - \gamma^{2} \right) e\cos \gamma (\phi - \phi_{0})
- \frac{3m_{10}G}{a(1 - e^{2})c^{2}} (1 + e\cos \gamma (\phi - \phi_{0}))^{2} = 0,
\end{equation}
If we neglect the last term
$- 3m_{10}Ga^{- 1}c^{- 2}(1 - e^{2})^{- 1}(1 + e\cos \gamma (\phi - \phi_{0}))^{2}$
in the left-hand side of the equality (\ref{4.30}), we get two equalities
$m_{10}Ga(1 - e^{2})h^{- 2}c^{- 2} = 1$, $\gamma = 1$ (see Ref. 9, Chap. 40, relation (40.2)).
The advance of Mercury's perihelion, observed from the Sun, is
$(\gamma^{- 1} - 1)\cdot 360\cdot 415 \cdot 3600^{''} = 0^{''}$ per century
($415$ "periods" of Mercury). By making use of the identity
$2\cos^{2} \gamma (\phi - \phi_{0}) = 1 + \cos 2\gamma (\phi - \phi_{0})$ we rewrite the
equality (\ref{4.30})
$$
1 - \frac{m_{10}Ga(1 - e^{2})}{h^{2}c^{2}} - \frac{3m_{10}G(2 + e^{2})}{2a(1 - e^{2})c^{2}}
$$
\begin{equation}
\label{4.19}
+  \left( 1 - \frac{6m_{10}G}{a(1 - e^{2})c^{2}} - \gamma^{2} \right) e\cos \gamma (\phi - \phi_{0})
- \frac{3m_{10}Ge^{2}}{2a(1 - e^{2})c^{2}} \cos 2\gamma (\phi - \phi_{0}) = 0.
\end{equation}
If we neglect the last term
$- 3m_{10}Ga^{- 1}c^{- 2}2^{- 1}e^{2}(1 - e^{2})^{- 1}\cos 2\gamma (\phi - \phi_{0})$
in the left-hand side of the equality (\ref{4.19}), we get two equalities:
$$
\frac{m_{10}Ga(1 - e^{2})}{h^{2}c^{2}} \, +
\frac{3m_{10}G(2 + e^{2})}{2a(1 - e^{2})c^{2}} \, = 1,
$$
\begin{equation}
\label{1.8} \gamma = \left( 1 - \frac{6m_{10}G}{a(1 - e^{2})c^{2}} \right)^{1/2}
\approx 1 - \frac{3m_{10}G}{a(1 - e^{2})c^{2}}.
\end{equation}
The equality $m_{10}Ga(1 - e^{2})h^{- 2}c^{- 2} = 1$ and the second equality (\ref{1.8})
are used in Ref. 9, Chap. 40, relations (40.5), (40.6). For the Mercury orbit the ellipse
major "semi - axis" $a = 0.5791\cdot 10^{11}m$ and the ellipse eccentricity $e = 0.21$.
The relations (\ref{1.5}), (\ref{1.10}), (\ref{1.8}) imply that the advance of Mercury's
perihelion observed from the Sun is
$(\gamma^{- 1} - 1)\cdot 360\cdot 415 \cdot 3600^{''} \approx
3m_{10}Ga^{- 1}c^{- 2}(1 - e^{2})^{- 1}\cdot 360\cdot 415 \cdot 3600^{''} \approx
3\omega^{2}a^{2} c^{- 2}(1 - e^{2})^{- 1} \cdot 360\cdot 415 \cdot
3600^{''} \approx 43^{''}.05 = 6\cdot 7^{''}.175$ (see the relations (\ref{1.6}) and
(\ref{1.8})) per century ($415$ "periods" of Mercury).

Misner, Thorne and Wheeler discuss and modify (Ref. 5, Chap. 40, Sec. 40.1, relation (40.1))
the Schwarzschild's metric (\ref{4.17})
$$
\sum_{\mu, \nu \, = \, 0}^{3} g_{\mu \nu} (x) dx^{\mu} dx^{\nu} =
\left( 1 - 2\frac{m_{10}G}{rc^{2}}  +
2 \left( \frac{m_{10}G}{rc^{2}} \right)^{2} \right) (dx^{0})^{2}
$$
\begin{equation}
\label{1.7} - \left( 1 + 2 \frac{m_{10}G}{rc^{2}} \right)
((dx^{1})^{2} + (dx^{2})^{2} + (dx^{3})^{2}), \, \,
r = ((x^{1})^{2} + (x^{2})^{2} + (x^{3})^{2})^{1/2}.
\end{equation}
A planet orbit is a solution of the geodesic equation (Ref. 5, Chap. 13, Sec. 13.4,
equation (13.36)) for the metric (\ref{1.7}). However, the approximation of Hamilton-Jacobi
equation is used for a planet orbit in (Ref. 5, Chap.40, Sec. 40.5). A planet orbit (Ref. 5,
Chap. 40, Sec. 40.5, equations (40.17), (40.18)) is given by the equation (\ref{1.5}) for
the precession coefficient (\ref{1.8}). The sketch of the equation (\ref{1.5}), (\ref{1.8})
proof is given in Exercise 40.4 from (Ref. 5, Chap. 40, Sec. 40.5). There is no a time
dependence of the orbit radus $r(t)$ and of the orbit angle $\phi (t)$ in (Ref. 5, Chap. 40,
Sec. 40.5). It needs the complete orbits of Mercury and the Earth to get the advance of
Mercury's perihelion, observed from the Earth.

In the Section III we solve the Kepler problem with the proper time $\tau$
\begin{equation}
\label{1.12}
\frac{d^{2}x^{i}}{d\tau^{2}}  = - \frac{m_{10}Gx^{i}}{r^{3}},
\, \, \frac{d}{d\tau} = \frac{dt}{d\tau} \frac{d}{dt}, \, \, i = 1,2,3,
\end{equation}
$$
\frac{dt}{d\tau} =  c\left( \sum_{\mu, \nu \, = \, 0}^{3}
g_{\mu \nu} (x(t)) \frac{dx^{\mu}}{dt} \frac{dx^{\nu}}{dt} \right)^{- 1/2}.
$$
The second relation (\ref{1.12}) is the relation (13.35) from (Ref. 5, Chap. 13, Sec. 13.4)
for the metric (\ref{1.7}). The exact solution of the Kepler problem (\ref{1.12}) is the
planet orbit (\ref{1.11}), (\ref{1.5}) with the precession coefficient $\gamma = 1$. It is
the approximate solution of the geodesic equations (Ref. 5, Chap. 13, Sec. 13.4, equations
(13.36)) for the metric (\ref{1.7}). It is possible to prove that the solution (\ref{1.11})
- (\ref{1.6}) of the equation (\ref{1.13}) is an approximate solution of the equation
(\ref{1.12}) for the world line $x^{\mu} (t)$, $x^{0} (t) = ct$.

\section{II. RELATIVISTIC PLANET ORBITS}
\setcounter{equation}{0}

Let us consider the relativistic Newton second law
\begin{equation}
\label{1.17} mc\frac{dt}{ds} \frac{d}{dt} \left( \frac{dt}{ds}
\frac{dx^{\mu}}{dt} \right) + qc^{- 1} \sum_{k\, =\, 0}^{N}
\, \sum_{\alpha_{1}, ..., \alpha_{k} \, =\, 0}^{3} \eta^{\mu \mu}
F_{\mu \alpha_{1} \cdots \alpha_{k}}(x)\frac{dt}{ds}
\frac{dx^{\alpha_{1}}}{0dt} \cdots \frac{dt}{ds}
\frac{dx^{\alpha_{k}}}{dt} = 0,
\end{equation}
$$
\frac{dt}{ds} = c^{- 1}\left( 1 - c^{- 2}|{\bf v}|^{2}\right)^{- 1/2},
\, \, v^{i} = \frac{dx^{i}}{dt}, \, \, i = 1,2,3.
$$
where $\mu = 0,...,3$ and the world line $x^{\mu}(t)$
satisfies the condition: $x^{0}(t)= ct$. The force is the polynomial
of the speed in the equation (\ref{1.17}). For an infinite series of the
speed we need to define the series convergence. The second relation (\ref{1.17})
implies the identities
\begin{equation}
\label{1.19} \sum_{\alpha \, =\, 0}^{3} \eta_{\alpha \alpha} \left(
\frac{dt}{ds} \frac{dx^{\alpha}}{dt} \right)^{2} = 1, \, \,
\sum_{\alpha \, =\, 0}^{3} \eta_{\alpha \alpha} \frac{dt}{ds}
\frac{dx^{\alpha}}{dt} \frac{dt}{ds} \frac{d}{dt} \left(
\frac{dt}{ds} \frac{dx^{\alpha}}{dt} \right) = 0.
\end{equation}
The equation (\ref{1.17}) and the second identity (\ref{1.19}) imply
\begin{equation}
\label{1.20} \sum_{k\, =\, 0}^{N} \sum_{\alpha_{1}, ..., \alpha_{k +
1} \, =\, 0}^{3} F_{\alpha_{1} \cdots \alpha_{k + 1}}(x)
\frac{dt}{ds} \frac{dx^{\alpha_{1}}}{dt} \cdots \frac{dt}{ds}
\frac{dx^{\alpha_{k + 1}}}{dt} = 0.
\end{equation}
Let the functions $F_{\alpha_{1} \cdots \alpha_{k + 1}}(x)$ satisfy
the equation (\ref{1.20}). Then three equations (\ref{1.17}) for
$\mu = 1,2,3$ are independent
\begin{eqnarray}
\label{1.21} m\frac{d}{dt} \left( (1 - c^{-
2}|{\bf v}|^{2})^{- 1/2}v^{i}\right) = \sum_{k\, =\, 0}^{N}
\, \, \sum_{\alpha_{1}, ..., \alpha_{k} \, =\, 0}^{3} \nonumber \\
qc^{- 1}(c^{2} - |{\bf v}|^{2})^{- (k - 1)/2} F_{i\alpha_{1} \cdots
\alpha_{k}}(x)\frac{dx^{\alpha_{1}}}{dt} \cdots
\frac{dx^{\alpha_{k}}}{dt},\, \, i = 1,2,3.
\end{eqnarray}

\noindent {\bf Lemma} (Ref. 3): {\it Let there exist Lagrange function}
$L({\bf x},{\bf v},t)$ {\it such that for any world line}
$x^{\mu}(t)$, $x^{0}(t) = ct$, {\it the relation}
\begin{eqnarray}
\label{1.22} \frac{d}{dt} \frac{\partial L}{\partial v^{i}} -
\frac{\partial L}{\partial x^{i}} = m\frac{d}{dt} \left( (1 - c^{-
2}|{\bf v}|^{2})^{- 1/2}v^{i}\right) - \sum_{k\, =\, 0}^{N}
\, \, \sum_{\alpha_{1}, ..., \alpha_{k} \, =\, 0}^{3} \nonumber \\
qc^{- 1}(c^{2} - |{\bf v}|^{2})^{- (k - 1)/2} F_{i\alpha_{1} \cdots
\alpha_{k}}(x)\frac{dx^{\alpha_{1}}}{dt} \cdots
\frac{dx^{\alpha_{k}}}{dt},\, \, i = 1,2,3,
\end{eqnarray}
{\it holds. Then Lagrange function has the form}
\begin{equation}
\label{1.23} L({\bf x},{\bf v},t) = - mc^{2}(1 - c^{- 2}|{\bf
v}|^{2})^{1/2} + q \sum_{i\, =\, 1}^{3} A_{i}({\bf
x},t)c^{- 1}v^{i} + qA_{0}({\bf x},t)
\end{equation}
{\it and the coefficients} $F_{i\alpha_{1} \cdots
\alpha_{k}}(x)$ {\it in the equations} (\ref{1.21}) {\it are}
\begin{equation}
\label{1.24} F_{i\alpha_{1} \cdots \alpha_{k}}(x) = 0,\, \, k \neq
1,\, i = 1,2,3,\, \alpha_{1}, ...,\alpha_{k} = 0,...,3,
\end{equation}
\begin{eqnarray}
\label{1.25} F_{ij}(x) = \frac{\partial A_{j}({\bf x},t)}{\partial
x^{i}} - \frac{\partial A_{i}({\bf x},t)}{\partial x^{j}},\, \, i,j
= 1,2,3, \nonumber
\\ F_{i0}(x) = \frac{\partial A_{0}({\bf x},t)}{\partial
x^{i}} - \frac{1}{c} \frac{\partial A_{i}({\bf x},t)}{\partial t},\,
\, i = 1,2,3.
\end{eqnarray}
We define the coefficients
\begin{equation}
\label{1.26} F_{00} (x) = 0,\, \, F_{0i} (x) = - F_{i0} (x),\, \, i = 1,2,3.
\end{equation}
Then the tensor $F_{\alpha \beta}(x)$ is antisymmetric and the identity
\begin{equation}
\label{1.27} \sum_{\alpha, \beta \, =\, 0}^{3} F_{\alpha
\beta}(x)\frac{dt}{ds} \frac{dx^{\alpha}}{dt} \frac{dt}{ds}
\frac{dx^{\beta}}{dt} = 0
\end{equation}
of the type (\ref{1.20}) holds. By making use of the
second identity (\ref{1.19}) and the relations (\ref{1.25}) -
(\ref{1.27}) we can rewrite the equation (\ref{1.21}) with the
coefficients (\ref{1.24}), (\ref{1.25}) as the relativistic Newton
second law with Lorentz force
\begin{eqnarray}
\label{1.28} mc\frac{dt}{ds} \frac{d}{dt} \left( \frac{dt}{ds}
\frac{dx^{\mu}}{dt} \right) = - q \eta^{\mu \mu} \sum_{\nu
\, =\, 0}^{3} F_{\mu \nu}(x)c^{- 1} \frac{dt}{ds} \frac{dx^{\nu}}{dt},
\nonumber \\ F_{\mu \nu}(x) = \frac{\partial A_{\nu}({\bf
x},t)}{\partial x^{\mu}} - \frac{\partial A_{\mu}({\bf
x},t)}{\partial x^{\nu}}, \, \, \mu, \nu = 0,...,3.
\end{eqnarray}
The interaction is defined by the product of the charge $q$ and the
external vector potential $A_{\mu}({\bf x},t)$.

Let a distribution $e_{0} (x) \in S^{\prime} ({\bf R}^{4})$ with
support in the closed upper light cone be a fundamental solution of
the wave equation
\begin{equation}
\label{112.33} - (\partial_{x},\partial_{x})
e_{0} (x) = \delta (x), \, \, (\partial_{x}, \partial_{x}) =
\left( \frac{\partial}{\partial x^{0}} \right)^{2} - \sum_{i\, =\, 1}^{3}
\left( \frac{\partial}{\partial x^{i}} \right)^{2}.
\end{equation}
The equation (\ref{112.33}) solution is unique in the class of
distributions with supports in the closed upper light cone
(Ref. 10, relation (2.42)). Due to (Ref. 11, Sect. 30) this unique causal
distribution is
\begin{equation}
\label{112.34} e_{0}(x) = -\, (2\pi)^{- 1} \theta (x^{0})\delta
((x,x)),
\end{equation}
$$
(x,y) = x^{0}y^{0} - \sum_{k\, =\, 1}^{3} x^{k}y^{k}, \, \,
\theta (x) = \left\{ {1, \hskip 0,5cm x \geq 0,} \atop
{0, \hskip 0,5cm x < 0.} \right.
$$
The relativistic causal Coulomb law is given by the equations of the type
(\ref{1.28})
\begin{eqnarray}
\label{12.1} m_{k} \frac{d}{dt} \left( \left( 1 - \frac{1}{c^{2}} \Bigl|
\frac{d{\bf x}_{k}}{dt}\Bigr|^{2} \right)^{- 1/2}
\frac{dx_{k}^{\mu}}{dt} \right) = - q_{k} \eta^{\mu \mu}
\sum_{\nu \, =\, 0}^{3} c^{- 1}\frac{dx_{k}^{\nu}}{dt}
\sum_{j\, =\, 1,2,\, j \neq k} F_{j;\mu \nu}(x_{k},x_{j})
\end{eqnarray}
where the strength $F_{j;\mu \nu}(x_{k},x_{j})$ is given by the relation
(\ref{1.2}) with Li\'enard - Wiechert vector potential of the type (\ref{1.3})
$$
A_{j; \mu}(x_{k},x_{j}) = - \, 4\pi q_{j}K \sum_{\nu \,
= \, 0}^{3} \eta_{\mu \nu} \int dt e_{0}(x_{k} - x_{j}(t))
\frac{dx_{j}^{\nu }(t)}{dt} =
$$
\begin{equation}
\label{12.4}
- \, q_{j}K \eta_{\mu \mu} \left( \frac{d}{dt}
x_{j}^{\mu}(t)\right) \left( c|{\bf x}_{k} - {\bf x}_{j}(t)| -
\sum_{i\, =\, 1}^{3} (x_{k}^{i} - x_{j}^{i}(t)) \frac{d}{dt}
x_{j}^{i}(t)\right)^{- 1} \Biggl|_{t = t(0)},
\end{equation}
\begin{equation}
\label{12.41}
x_{k}^{0} - ct(0) = |{\bf x}_{k} - {\bf x}_{j}(t(0))|.
\end{equation}
Here $K$ is the constant of the causal electromagnetic interaction
for two particles with the charges $q_{j}$. The support of the
distribution (\ref{112.34}) lies in the upper light cone boundary.
The interaction speed is equal to that of light. It is easy to prove
the second relation (\ref{12.4}) by making change of the integration
variable
\begin{equation}
\label{12.17} x_{k}^{0} - ct(s) = (|{\bf x}_{k} - {\bf
x}_{j}(t(s))|^{2} + s)^{1/2}.
\end{equation}
For $s = 0$ the relation (\ref{12.17}) coincides with the relation
(\ref{12.41}).

The equations (\ref{12.1}), (\ref{1.2}), (\ref{12.4}) are the relativistic
causal version of the Coulomb law. The Lorentz invariant distribution
(\ref{112.34}) defines the delay. The Lorentz invariant solutions of the
equation (\ref{112.33}) are described in Ref. 3. By making use
of these solutions it is possible to describe the Lorentz covariant
equations of the type (\ref{12.1}), (\ref{1.2}), (\ref{12.4}). The
equations (\ref{12.1}), (\ref{1.2}), (\ref{12.4}) are Lorentz covariant
and causal due to the distribution (\ref{112.34}). The quantum version
of the equations (\ref{12.1}), (\ref{1.2}), (\ref{12.4}) is defined in
Ref. 10. The solutions of these causal equations do not contain the
diverging integrals similar to the diverging integrals of the quantum
electrodynamics.

For a world line $x_{j}^{\mu}(t)$ we define the vector
$$
J^{\mu}(x,x_{j}) =  - (\partial_{x},\partial_{x})
\int dt e_{0}(x - x_{j}(t)) \frac{dx_{j}^{\mu }(t)}{dt} =
\int dt \delta (x - x_{j}(t)) \frac{dx_{j}^{\mu }(t)}{dt} =
$$
\begin{equation}
\label{12.10}  \left( \frac{d}{dx^{0}}
x_{j}^{\mu} \left( c^{- 1}x^{0} \right) \right) \delta \left( {\bf
x} - {\bf x}_{j} \left( c^{- 1} x^{0} \right) \right), \mu = 0,...,3.
\end{equation}
The condition $x_{j}^{0}(t) = ct$ implies the equalities
\begin{equation}
\label{12.121} \frac{\partial}{\partial x^{0}} J^{0}(x,x_{j}) = -
\sum_{i\, =\, 1}^{3} \left( \frac{d}{dx^{0}} x_{j}^{i} \left( c^{-
1}x^{0} \right) \right) \frac{\partial}{\partial x^{i}} \delta
\left( {\bf x} - {\bf
x}_{j} \left( c^{- 1} x_{k}^{0} \right) \right),
\end{equation}
\begin{equation}
\label{12.122}
\frac{\partial}{\partial x^{i}} J^{i}(x,x_{j}) = \left(
\frac{d}{dx^{0}} x_{j}^{i} \left( c^{- 1}x^{0} \right) \right)
\frac{\partial}{\partial x^{i}} \delta \left( {\bf x} - {\bf x}_{j}
\left( c^{- 1} x^{0} \right) \right),\, i = 1,2,3.
\end{equation}
The equalities (\ref{12.121}), (\ref{12.122}) imply the continuity equation
\begin{equation}
\label{12.12} \sum_{\mu \, =\, 0}^{3} \frac{\partial}{\partial
x^{\mu}} J^{\mu}(x,x_{j}) = 0.
\end{equation}
The integration of the relation
\begin{equation}
\label{12.13} e_{0}(x - x_{j}(t)) \frac{dx_{j}^{\mu }(t)}{dt} = \int
d^{4}y e_{0}(x - y) \delta (y - x_{j}(t))
\frac{dx_{j}^{\mu}(t)}{dt}
\end{equation}
yields
\begin{equation}
\label{12.14} \int dt e_{0}(x - x_{j}(t)) \frac{dx_{j}^{\mu
}(t)}{dt} = \int d^{4}y e_{0}(x - y) J^{\mu}(y,x_{j}).
\end{equation}
The relations (\ref{12.12}), (\ref{12.14}) imply the gauge condition
for the vector potential (\ref{12.4})
\begin{equation}
\label{12.16} \sum_{\mu \, =\, 0}^{3} \eta^{\mu \mu}
\frac{\partial}{\partial x^{\mu}} A_{j;\mu}(x,x_{j}) = 0.
\end{equation}
Due to the gauge condition (\ref{12.16}) the tensor (\ref{1.2}),
(\ref{12.4}) satisfies Maxwell equations with the current
proportional to the current (\ref{12.10}).

The substitution $K = - \, G$ and two positive or two negative
gravitational masses $q_{1} = \pm \, m_{1}$, $q_{2} = \pm \, m_{2}$
into the equations (\ref{12.1}), (\ref{1.2}), (\ref{12.4})
yields the relativistic causal Newton gravity law (\ref{1.1}) -
(\ref{1.3}). By changing the constants $K = - \, G$, $q_{1} = \pm \, m_{1}$,
$q_{2} = \pm \, m_{2}$ in the equations from Ref. 10 we have the quantum
version of the equations (\ref{1.1}) - (\ref{1.3}). The substitution
$K = - \, G$ and also one positive and one negative gravitational masses
$q_{1} = \pm \, m_{1}$, $q_{2} = \mp \, m_{2}$ into the equations (\ref{12.1}),
(\ref{1.2}), (\ref{12.4}) yields the galaxies scattering with an acceleration.
For the negative constant $K$ of the causal electromagnetic interaction the
protons and electeons really couldn't exist together.

The idea of the electromagnetic and gravitational interactions similarity
is not new. Einstein${}^{12}$: "The theoretical physicists studying the problems
of general relativity can hardly doubt now that the gravitational and
electromagnetic fields should have the same nature."

The relativistic Newton gravity law for the solar system was proposed in Ref. 3
$$
\frac{d}{dt} \left( \left( 1 - \frac{1}{c^{2}} \Bigl| \frac{d{\bf
x}_{k}}{dt}\Bigr|^{2} \right)^{- 1/2} \frac{dx_{k}^{\mu}}{dt}
\right) =
$$
\begin{equation}
\label{2.1} - m_{10}\eta^{\mu \mu} \sum_{\nu \, =\, 0}^{3}
\frac{1}{c} \frac{dx_{k}^{\nu}}{dt} \sum_{j\, =\, 1,...,10,\, j \neq k}
\frac{1}{m_{10}} \, F_{j;\, \mu \nu}(x_{k},x_{j}),
\end{equation}
$k = 1,...,10$, $\mu = 0,...,3$. We give the number $k = 1$ for Mercury, the
number $k = 2$ for Venus, the number $k = 3$ for the Earth, the number $k = 4$ for
Mars, the number $k = 5$ for Jupiter, the number $k = 6$ for Saturn, the number
$k = 7$ for Uranus, the number $k = 8$ for Neptune, the number $k = 9$ for Pluto
and the number $k = 10$ for the Sun.

The calculation of the equation (\ref{2.1}) orbits is a difficult and laborious task.
We consider the Sun resting at the coordinates origin. Substituting the Sun world
line $x_{10}^{0}(t) = ct$, $x_{10}^{i}(t) = 0$, $i = 1,2,3$, into the equalities
(\ref{1.2}), (\ref{1.3}) we have
\begin{equation}
\label{2.8} F_{10;ij}(x;x_{10}) = 0,\, \, i,j = 1,2,3,\, \,
F_{10;i0}(x;x_{10}) = - m_{10}G|{\bf x}|^{- 3}x^{i},\, \, i = 1,2,3.
\end{equation}
Due to the relations (\ref{1.2}), (\ref{1.3}) the value $ m_{10}^{- 1}F_{j;\, \mu
\nu}(x_{k},x_{j})$ is proportional to $m_{j}m_{10}^{- 1}$, $j = 1,...,9$.
According to (Ref. 5, Chap. 25, Sec. 25.1, Appendix 25.1) for Jupiter the ratio
$m_{5}m_{10}^{- 1} \approx 0.95 \cdot 10^{- 3}$ is maximal. For the Earth the
ratio $m_{3}m_{10}^{- 1} \approx 3.01 \cdot 10^{- 6}$. We neglect the action of
any planet on all of the other planets. Substituting the Sun world line
$x_{10}^{0}(t) = ct$, $x_{10}^{i}(t) = 0$, $i = 1,2,3$, and the relations (\ref{2.8})
into the equations (\ref{2.1}) we get
\begin{equation}
\label{2.9} \frac{d}{dt} \left( \left( 1 - \frac{1}{c^{2}} \Bigl| \frac{d{\bf
x}_{k}}{dt}\Bigr|^{2}\right)^{- 1/2} \frac{d{\bf x}_{k}}{dt} \right) =
- \frac{m_{10}G}{|{\bf x}_{k}|^{3}} \, {\bf x}_{k},
\end{equation}
$k = 1,...,9$. Due to the equations (\ref{2.9}) the angular momentum and the energy
\begin{equation}
\label{2.10} M_{l}({\bf x}_{k}) = \sum_{i,j = 1}^{3} \epsilon_{ijl}
\left( x_{k}^{i}\frac{dx_{k}^{j}}{dt} -
x_{k}^{j}\frac{dx_{k}^{i}}{dt} \right) \left( 1 - \frac{1}{c^{2}}
\Biggl| \frac{d{\bf x}_{k}}{dt} \Biggr|^{2} \right)^{- 1/2},
\end{equation}
\begin{equation}
\label{2.11} E({\bf x}_{k}) = c^{2}\left( 1 - \frac{1}{c^{2}} \Biggl|
\frac{d{\bf x}_{k}}{dt}\Biggr|^{2} \right)^{- 1/2} - \frac{m_{10}G}{|{\bf x}_{k}|} \, ,
\end{equation}
$l = 1,2,3$, $k = 1,...,9$, are time independent. The antisymmetric
in all indices tensor $\epsilon_{ijl}$ has the normalization $\epsilon_{123} = 1$.
Let the third axis coincide with the constant vector (\ref{2.10}). The vector
${\bf x}_{k}$ is orthogonal to the constant vector (\ref{2.10}). We introduce the
polar coordinates in the plane orthogonal to the vector (\ref{2.10})
\begin{equation}
\label{2.12} x_{k}^{1} (t) = r_{k}(t)\cos \phi_{k} (t),\, \,
x_{k}^{2} (t) = r_{k}(t)\sin \phi_{k} (t), \, \, x_{k}^{3} (t) =  0,
\end{equation}
$k = 1,...,9$. The relations (\ref{2.10}), (\ref{2.11}) imply
$$
r_{k}^{2}(t)\left( E({\bf x}_{k}) + \frac{m_{10}G}{r_{k}(t)} \right)
\frac{d\phi_{k}}{dt} = c^{2}|{\bf M}({\bf x}_{k})|,
$$
$$
\left( E({\bf x}_{k}) + \frac{m_{10}G}{r_{k}(t)} \right)^{2}
\left( \frac{dr_{k}}{dt} \right)^{2} =
$$
\begin{equation}
\label{2.121} c^{2} ((E({\bf x}_{k}))^{2} - c^{4})
+ \frac{2m_{10}Gc^{2}E({\bf x}_{k})}{r_{k}(t)} +
\frac{m_{10}^{2}G^{2}c^{2} - |{\bf M}({\bf x}_{k})|^{2}c^{4}}{r_{k}^{2}(t)},
\end{equation}
$k = 1,...,9$. Let the constants (\ref{2.10}), (\ref{2.11}) satisfy the inequalities
\begin{equation}
\label{2.18} (E({\bf x}_{k}))^{2} - c^{4} < 0,
\end{equation}
\begin{equation}
\label{2.13} c^{2}|{\bf M}({\bf x}_{k})|^{2} - m_{10}^{2}G^{2} > 0,
\end{equation}
\begin{equation}
\label{2.14} c^{2}|{\bf M}({\bf x}_{k})|^{2}((E({\bf x}_{k}))^{2} - c^{4}) +
m_{10}^{2}G^{2}c^{4} > 0,
\end{equation}
$k = 1,...,9$. Due to Ref. 3 the equations (\ref{2.121}) have the solutions
\begin{equation}
\label{2.15} \frac{a_{k}(1 - e_{k}^{2}) }{r_{k}(t)} = 1 + e_{k} \cos
\gamma_{k} (\phi_{k} (t) - \phi_{k;0} ).
\end{equation}
The orbit radius $r_{k}(t)$ is given by
$$
r_{k}(t(\xi_{k})) = m_{10}GE({\bf x}_{k})(c^{4} - (E({\bf
x}_{k}))^{2})^{- 1} (1 + e_{k}\sin \xi_{k}),
$$
\begin{equation}
\label{2.19}
t(\xi_{k}) = m_{10}Gc^{- 1}(c^{4} - (E({\bf
x}_{k}))^{2})^{- 3/2} (c^{4}(\xi_{k} -
\xi_{k;0}) - e_{k}(E({\bf x}_{k}))^{2}\cos \xi_{k}).
\end{equation}
The perihelion angle $\phi_{k;0}$, the parameter $\xi_{k;0}$ and the values
$$
a_{k}(1 - e_{k}^{2}) = (c^{2}|{\bf M}({\bf x}_{k})|^{2} -
m_{10}^{2}G^{2})(m_{10}GE({\bf x}_{k}))^{- 1},
$$
$$
e_{k} = (c^{2}|{\bf M}({\bf x}_{k})|^{2}((E({\bf
x}_{k}))^{2} - c^{4}) + m_{10}^{2}G^{2}c^{4})^{1/2} (m_{10}GE({\bf
x}_{k}))^{- 1},
$$
\begin{equation}
\label{2.16}
\gamma_{k} = \left( 1 - \frac{m_{10}^{2}G^{2}}{c^{2}|{\bf M}({\bf x}_{k})|^{2}}
\right)^{1/2},\, \, k = 1,...,9,
\end{equation}
are constant. The orbit eccentricities: $e_{1} = 0.21$, $e_{2} = 0.007$,
$e_{3} = 0.017$, $e_{4} = 0.093$, $e_{5} = 0.048$, $e_{6} = 0.056$,
$e_{7} = 0.047$, $e_{8} = 0.009$, $e_{9} = 0.249$. Therefore $0 < e_{k} < 1$,
$k = 1,...,9$. Let us suppose $E({\bf x}_{k}) > 0$, $k = 1,...,9$. The
inequalities (\ref{2.13}) and $e_{k}^{2} < 1$ imply the inequality (\ref{2.18}).
For $e_{k}^{2} < 1$ the equation (\ref{2.15}) defines an ellipse with a
precession given by the coefficient $\gamma_{k}$. The focus of this ellipse is
the coordinates origin. It is a relativistic analogue of the first Kepler law.
The time independence of the vector (\ref{2.10}) is the relativistic second Kepler law.
The equations (\ref{2.19}) define the time dependence of the radius $r_{k}$. For
$c \rightarrow \infty$ the equations (\ref{2.9}) solutions tend to the Kepler
problem solutions. (For the Kepler problem equations the multiplier
$( 1 - c^{- 2}| d{\bf x}_{k}/dt|^{2})^{- 1/2}$ is absent in the equations
(\ref{2.9})).

Let us express the constants in the equations (\ref{2.15}), (\ref{2.19}) trough
the astronomical data. The ellipse (\ref{2.15}) major "semi - axis" is
equal to
\begin{equation}
\label{2.20} a_{k} = m_{10}GE({\bf x}_{k})(c^{4} - (E({\bf x}_{k}))^{2})^{- 1},
\, \, k = 1,...,9.
\end{equation}
For the parameters $\pm \pi /2$ we have the extremal radii
\begin{equation}
\label{2.22} r_{k}(t(\pm \pi /2)) =  a_{k}(1 \pm e_{k}),\, \, k = 1,...,9.
\end{equation}
Hence, the "period" of the motion along the ellipse (\ref{2.15}) is
equal to
\begin{equation}
\label{2.24} T_{k} = 2(t_{k}(\pi /2) - t_{k}(- \pi /2)) = 2\pi
m_{10}Gc^{3}(c^{4} - (E({\bf x}_{k}))^{2})^{- 3/2}, \, \, k =
1,...,9.
\end{equation}
Let us define the mean "angular frequency" $\omega_{k} = 2\pi
T_{k}^{- 1}$. The relation (\ref{2.24}) implies
\begin{eqnarray}
\label{2.26} \omega_{k} = (c^{4} - (E({\bf
x}_{k}))^{2})^{3/2}(m_{10}Gc^{3})^{- 1},
\nonumber \\
(E({\bf x}_{k}))^{2} = c^{2}(c^{2} - (\omega_{k} m_{10}G)^{2/3}),\,
\, k = 1,...,9.
\end{eqnarray}
The substitution of the expression (\ref{2.26}) into the equality
(\ref{2.20}) yields
\begin{equation}
\label{2.27} m_{10}G = \omega_{k}^{2} a_{k}^{3} \left( 2^{- 1}(1 +
\sigma_{k} (1 - (2a_{k}\omega_{k} c^{- 1})^{2})^{1/2})\right)^{-
3/2}, \, \, \sigma_{k} = \pm 1, \, \, k = 1,...,9.
\end{equation}
According to (Ref. 5, Chap. 25, Sec. 25.1, Appendix 25.1) the
values $\omega_{k}^{2} a_{k}^{3}c^{- 2} = 1477m$ for $k = 1,2,3,4,6$
(Mercury, Venus, the Earth, Mars and Saturn), the values
$\omega_{l}^{2} a_{l}^{3}c^{- 2} = 1478m$ for $l = 5,8$ (Jupiter and Neptune),
the value $\omega_{7}^{2} a_{7}^{3}c^{- 2} = 1476m$ for Uranus, the value
$\omega_{9}^{2} a_{9}^{3}c^{- 2} = 1469m$ for Pluto; the major semi-axes
$a_{1} = 0.5791\cdot 10^{11}m$, $a_{2} = 1.0821\cdot 10^{11}m$,
$a_{3} = 1.4960\cdot 10^{11}m$, $a_{4} = 2.2794\cdot 10^{11}m$,
$a_{5} = 7.783\cdot 10^{11}m$, $a_{6} = 14.27\cdot 10^{11}m$,
$a_{7} = 28.69\cdot 10^{11}m$, $a_{8} = 44.98\cdot 10^{11}m$,
$a_{9} = 59.00\cdot 10^{11}m$. The values
$\omega_{k}^{2}a_{k}^{2} c^{- 2} = a_{k}^{- 1}\cdot \omega_{k}^{2}a_{k}^{3} c^{- 2}$,
$k = 1,...,9$, are negligible and therefore $m_{10}G \approx \omega_{k}^{2}
a_{k}^{3}(2^{- 1}(1 + \sigma_{k} ))^{- 3/2}$. For $\sigma_{k} = 1$
this expression agrees with the third Kepler law (\ref{1.10}): $m_{10}G =
\omega_{k}^{2} a_{k}^{3}$. Choosing $\sigma_{k} = 1$ in the relation
(\ref{2.27}) we get the third Kepler law
\begin{equation}
\label{2.28}  m_{10}G = \omega_{k}^{2} a_{k}^{3}\left( 2^{- 1}(1 +
(1 - 4\omega_{k}^{2} a_{k}^{2} c^{- 2})^{1/2})\right)^{- 3/2}
\approx \omega_{k}^{2} a_{k}^{3}
\left( 1 + \frac{3}{2}  \omega_{k}^{2} a_{k}^{2}c^{- 2} \right)
\end{equation}
for the orbits (\ref{2.15}), (\ref{2.19}). The Sun mass
values (\ref{2.28}) obtained in the relativistic Kepler problem agrees perfectly
with values $\omega_{k}^{2} a_{k}^{3}$ obtained in Kepler problem. The
substitution of the vector (\ref{2.12}), $r_{k}(t) = a_{k}$,
$\phi_{k} (t) = \omega_{k} (t - t_{0})$, into the equation (\ref{2.9}) yields
the third Kepler law
\begin{equation}
\label{2.29}  m_{10}G = \omega_{k}^{2} a_{k}^{3}(1 - \omega_{k}^{2}a_{k}^{2}c^{- 2})^{- 1/2}
\approx \omega_{k}^{2} a_{k}^{3}\left( 1 + \frac{1}{2} \omega_{k}^{2}a_{k}^{2}c^{- 2}\right),
\, \, k = 1,...,9,
\end{equation}
for the equation (\ref{2.9}) periodic circular orbits. The substitution
of the expression (\ref{2.28}) into the equality (\ref{2.26}) yields
\begin{equation}
\label{2.30} c^{- 4}(E({\bf x}_{k}))^{2} = 1 - 2\omega_{k}^{2}
a_{k}^{2}c^{- 2}\left(1 + (1 - 4\omega_{k}^{2} a_{k}^{2} c^{-
2})^{1/2}\right)^{- 1} \approx 1 - \omega_{k}^{2} a_{k}^{2}c^{-
2},\, \, k = 1,...,9.
\end{equation}
By making use of the relations (\ref{2.16}), (\ref{2.20}), (\ref{2.28}),
(\ref{2.30}) we have
\begin{equation}
\label{2.31} \gamma_{k} =
\left( 1 + 4\omega_{k}^{2} a_{k}^{2} c^{- 2} (1 - e_{k}^{2})^{- 1}
\left(1 + (1 - 4\omega_{k}^{2} a_{k}^{2}c^{- 2})^{1/2}\right)^{- 2}
\right)^{- 1/2} \approx 1 - \frac{\omega_{k}^{2}
a_{k}^{2} c^{- 2}}{2(1 - e_{k}^{2})},
\end{equation}
$k = 1,...,9$. The value
$2^{- 1}\omega_{k}^{2}a_{k}^{2} c^{- 2}(1 - e_{k}^{2})^{- 1}$ is maximal for Mercury:
$2^{- 1}\omega_{1}^{2}a_{1}^{2} c^{- 2}(1 - e_{1}^{2})^{- 1} \approx 1.3341 \cdot 10^{- 8}$.
The precession coefficients (\ref{2.31}) of the orbits (\ref{2.15}) are practically equal
to one for all planets. It agrees with Tycho Brahe's astronomical observations used by Kepler.
The relations (\ref{2.15}), (\ref{2.31}) imply the perihelion angle
\begin{equation}
\label{2.32} \phi_{k;l} \approx \phi_{k;0} + 2\pi l(1 + 2^{- 1}\omega_{k}^{2}
a_{k}^{2} c^{- 2} (1 - e_{k}^{2})^{- 1}), \, \, l = 0,\pm 1,\pm 2,....
\end{equation}
The substitution of the relations (\ref{2.20}), (\ref{2.28}),
(\ref{2.30}), (\ref{2.31}) into the equalities (\ref{2.15}), (\ref{2.19}) yields
\begin{equation}
\label{2.33} e_{k}\cos \left( \left( 1 - 2^{- 1}\omega_{k}^{2}
a_{k}^{2} c^{- 2}(1 - e_{k}^{2})^{- 1} \right) (\phi_{k} (t) -
\phi_{k;0})\right) \approx a_{k}(1 - e_{k}^{2})r_{k}^{- 1} (t) - 1,
\end{equation}
\begin{equation}
\label{2.34} r_{k}(t_{k}(\xi_{k})) \approx a_{k}(1 + e_{k}\sin \xi_{k}), \, \,
\omega_{k} t_{k}(\xi_{k}) \approx \xi_{k} - \xi_{k;0} -
e_{k}(1 - \omega_{k}^{2} a_{k}^{2} c^{- 2})\cos \xi_{k},
\end{equation}
$k = 1,...,9$. Let us define the constant $\xi_{k;0}$ in the
second equality (\ref{2.34}) by choosing the initial time moment
$t_{k}(0) = 0$. Then the equalities (\ref{2.34}) have the form
\begin{equation}
\label{2.35}
r_{k}(t_{k}(\xi_{k})) \approx a_{k}(1 + e_{k}\sin \xi_{k}), \, \,
\omega_{k} t_{k}(\xi_{k}) \approx \xi_{k} + e_{k}(1 -
\omega_{k}^{2} a_{k}^{2} c^{- 2})\left( 1 - \cos \xi_{k} \right),
\end{equation}
$k = 1,...,9$. Let the direction of the first axis be orthogonal to the
vector ${\bf M}({\bf x}_{1})$. Let the direction of the third axis
coincide with the direction of vector ${\bf M}({\bf x}_{3})$. Then the
second axis lies in the plane stretched on the vectors ${\bf M}({\bf
x}_{1})$ and ${\bf M}({\bf x}_{3})$. Due to the relations
(\ref{2.12}) the Mercury and Earth orbits
$$
x_{1}^{1} (t) = r_{1}(t)\cos \phi_{1} (t),\, \, x_{1}^{2} (t) = -
r_{1}(t)\cos \theta_{1} \sin \phi_{1} (t),\, \, x_{1}^{3} (t) =
r_{1}(t)\sin \theta_{1} \sin \phi_{1} (t),
$$
\begin{equation}
\label{2.36} x_{3}^{1} (t) = r_{3}(t)\cos \phi_{3} (t),\, \,
x_{3}^{2} (t) = r_{3}(t)\sin \phi_{3} (t), \, \, x_{3}^{3} = 0
\end{equation}
where the inclination of Mercury orbit plane $\theta_{1} = 7^{o}$
and the values $r_{k}(t),\phi_{k} (t)$, $k = 1,3$, satisfy the
equations (\ref{2.33}), (\ref{2.35}). For the definition of Mercury
and the Earth trajectories it is necessary to define the perihelion
angles $\phi_{1;0}$, $\phi_{3;0}$ in the equations (\ref{2.33}).

"Observations of Mercury do not give the absolute position of the
planet in space but only the direction of a line from the planet to
the observer."  (Ref. 6, p. 363.) The advance of Mercury's
perihelion is given by the angle
\begin{eqnarray}
\label{2.37} \cos \alpha = \frac{({\bf x}_{1}(t_{1}(\xi_{1,1})) -
{\bf x}_{3}(t_{3}(\xi_{3,1})), {\bf x}_{1}(t_{1}(\xi_{1,2})) -
{\bf x}_{3}(t_{3}(\xi_{3,2})))}{|{\bf x}_{1}(t_{1}(\xi_{1,1})) -
{\bf x}_{3}(t_{3}(\xi_{3,1}))||{\bf x}_{1}(t_{1}(\xi_{1,2})) -
{\bf x}_{3}(t_{3}(\xi_{3,2}))|}, \nonumber \\ c(t_{3}(\xi_{3,k}) -
t_{1}(\xi_{1,k})) = |{\bf x}_{1}(t_{1}(\xi_{1,k})) - {\bf
x}_{3}(t_{3}(\xi_{3,k}))|,\, \, k = 1,2, \nonumber \\
t_{1}(\xi_{1,2}) - t_{1}(\xi_{1,1}) \leq 100T_{3} \leq
t_{1}(\xi_{1,2}) - t_{1}(\xi_{1,1}) + T_{1}
\end{eqnarray}
where the parameters $\xi_{1,1}$, $\xi_{1,2}$ are defined by
Mercury's perihelion points, the parameters $\xi_{3,1}$,
$\xi_{3,2}$ are the solutions of the second equation (\ref{2.37}),
the numbers $T_{1}$, $T_{3}$ are the orbit "periods" of
Mercury and the Earth. The quotient $T_{3}/T_{1}$ of the Earth and
Mercury orbit "periods" is approximately equal to $4.15$.

By making use of the equations (\ref{2.35}) we obtain the parameters
corresponding to Mercury's perihelion points:
$$
a_{1}^{- 1}r_{1}(t_{1}(\xi_{1,k})) \approx  1 - e_{1},\, \,
\omega_{1} t_{1}(\xi_{1,k}) \approx \pi \left( 2l_{k} +
3/2 \right) + e_{1}(1 - \omega_{1}^{2} a_{1}^{2} c^{- 2}),
$$
\begin{equation}
\label{2.39}
\xi_{1,k} \approx \pi \left( 2l_{k} + 3/2
\right),\, \, k = 1,2,
\end{equation}
where $l_{k}$  are the integers. The first relation (\ref{2.39})
coincides with the equality (\ref{2.22}).

According to (Ref. 5, Chap. 25, Sec. 25.1, Appendix 25.1)
$c^{- 1}\omega_{1} = 275.8\cdot 10^{- 17}m^{- 1}$, $c^{-
1}\omega_{3} = 66.41\cdot 10^{- 17}m^{- 1}$, $a_{1} = 0.5791\cdot
10^{11}m$, $a_{3} = 1.4960\cdot 10^{11}m$. The substitution of the second
equality (\ref{2.39}) into the third relation (\ref{2.37}) yields
$l_{2} - l_{1} = 415$.

Due to the second relation (\ref{2.37})
\begin{equation}
\label{2.40} {\bf x}_{3}(t_{3}(\xi_{3,k})) = {\bf
x}_{3}(t_{1}(\xi_{1,k})) + c^{- 1}|{\bf x}_{1}(t_{1}(\xi_{1,k})) -
{\bf x}_{3}(t_{3}(\xi_{3,k}))|{\bf v}_{3}(t_{3,k}^{\prime}),\, \, k
= 1,2.
\end{equation}
The Earth speed is small compared with the speed of light: $c^{-
1}|{\bf v}_{3}| \approx c^{- 1}\omega_{3} a_{3} \approx 0.9935\cdot
10^{- 4}$. We neglect this value ($\arcsin 10^{- 4} \approx
0^{o}.0057$). Then the relations (\ref{2.37}), (\ref{2.40}) imply
\begin{equation}
\label{2.41} \cos \alpha \approx \frac{({\bf
x}_{1}(t_{1}(\xi_{1,1})) - {\bf x}_{3}(t_{1}(\xi_{1,1})), {\bf
x}_{1}(t_{1}(\xi_{1,2})) - {\bf x}_{3}(t_{1}(\xi_{1,2})))}{|{\bf
x}_{1}(t_{1}(\xi_{1,1})) - {\bf x}_{3}(t_{1}(\xi_{1,1}))||{\bf
x}_{1}(t_{1}(\xi_{1,2})) - {\bf x}_{3}(t_{1}(\xi_{1,2}))|}
\end{equation}
where the parameters $\xi_{1,k}$, $k = 1,2$, are given by the third
relation (\ref{2.39}) and the relation $l_{2} = l_{1} + 415$.

Let us consider Mercury's perihelion points corresponding to the integers
$l_{1} = 0$ and $l_{2} = 415$. The substitution of the values corresponding
to the Mercury's perihelion, defined by the first equation (\ref{2.39}),
into the equation (\ref{2.33}) yields
$$
r_{1}\left( t_{1}(\pi \left( 2l + 3/2
\right) ) \right) \approx  a_{1}(1 - e_{1}),
$$
\begin{equation}
\label{2.42}
\phi_{1} \left( t_{1}\left( \pi \left( 2l + 3/2 \right) \right) \right)
\approx \phi_{1;0} + 2\pi l\left( 1 + 2^{- 1}\omega_{1}^{2} a_{1}^{2}
c^{- 2}(1 - e_{1}^{2})^{- 1}\right)
\end{equation}
since the value $\omega_{1}^{2}a_{1}^{2} c^{- 2} \approx 2.5509\cdot
10^{- 8}$ is negligible. We substitute the time, defined by the
second relation (\ref{2.39}), into the second relation (\ref{2.35}) for
the Earth
$$
r_{3}(t_{1}(\xi_{3} (l))) \approx a_{3}(1 + e_{3})\sin \xi_{3} (l),
$$
$$
\omega_{3} t_{1}\left( \pi \left( 2l +
3/2 \right) \right) \approx \omega_{3} \omega_{1}^{- 1}
\left( \pi \left( 2l + 3/2 \right) + e_{1}(1 -
\omega_{1}^{2} a_{1}^{2}
c^{- 2})\right) \approx
$$
\begin{equation}
\label{2.43}
\xi_{3} (l) - e_{3}(1 - \omega_{3}^{2} a_{3}^{2} c^{- 2})\left(
\cos \xi_{3} (l) - 1\right).
\end{equation}
Solving the second equation (\ref{2.43}) we get $\xi_{3} (0)
\approx 1.1748$, $\xi_{3} (415) \approx 629.09$. Substituting
these values in the first equation (\ref{2.43}) we have $a_{3}^{-
1}r_{3}(\xi_{3} (0)) \approx 1.0157$, $a_{3}^{- 1}r_{3}(\xi_{3} (415))
\approx 1.0118$. We substitute the first equation (\ref{2.43}) in the
equation (\ref{2.33}) for the Earth
\begin{equation}
\label{2.44} \cos \left( \left( 1 - \frac{\omega_{3}^{2} a_{3}^{2}
c^{- 2}}{2(1 - e_{3}^{2})}\right) \left( \phi_{3} \left( t_{1}
\left( \pi \left( 2l + 3/2 \right) \right) \right) -
\phi_{3;0} \right) \right) \approx - \frac{e_{3} + \sin \xi_{3}
(l)}{1 + e_{3}\sin \xi_{3} (l)}.
\end{equation}
The function in the right - hand side of the equation (\ref{2.44})
is monotonic with respect to the variable $e_{3}$ on the interval $0
\leq e_{3} \leq 1$. Calculating the values of this function at the
points $e_{3} = 0,1$ we get the estimation for the module of this
function which implies that the equation (\ref{2.44}) has a
solution. Substituting the solutions $\xi_{3} (l)$, $l =
0,415$, of the second equation (\ref{2.43}) in the
equation (\ref{2.44}) we get the angles in radians
$$
\phi_{3} \left( t_{1} \left( \pi \left( 2\cdot 0 + 3/2
\right) \right) \right) \approx \phi_{3;0} + 2.7521,
$$
\begin{equation}
\label{2.441} \phi_{3} \left( t_{1} \left( \pi \left( 2\cdot 415 +
3/2 \right) \right) \right) \approx \phi_{3;0} + 2.3544 +
2\pi \cdot 99 \left( 1 + 2^{- 1}\omega_{3}^{2} a_{3}^{2} c^{- 2}(1
- e_{3}^{2})^{- 1}\right)
\end{equation}
since the value $\omega_{3}^{2}a_{3}^{2} c^{- 2} \approx 0.9870\cdot
10^{- 8}$ is negligible. Substituting the radii and the angles
(\ref{2.42}), the radii (\ref{2.43}) and the angles (\ref{2.441}) in
the equations (\ref{2.36}), (\ref{2.41}) we get the equation
\begin{eqnarray}
\label{2.45}  \cos \alpha (0,415) \approx (((1 - e_{1})\cos
\phi_{1;0} - 1.0157a_{3}a_{1}^{- 1}\cos (\phi_{3;0} + 2.7521))
\nonumber
\\ \times ((1 - e_{1})\cos (\phi_{1;0} + 415\pi
\omega_{1}^{2} a_{1}^{2} c^{- 2}(1 - e_{1}^{2})^{- 1}) \nonumber
\\ - 1.0118a_{3}a_{1}^{- 1}\cos (\phi_{3;0} + 2.3544 + 99 \pi
\omega_{3}^{2} a_{3}^{2} c^{- 2}(1 - e_{3}^{2})^{- 1})) \nonumber \\
+ (0.99255(1 - e_{1})\sin \phi_{1;0} + 1.0157a_{3}a_{1}^{- 1}\sin
(\phi_{3;0} + 2.7521)) \nonumber
\\ \times ( 0.99255(1 - e_{1})\sin ( \phi_{1;0} + 415\pi \omega_{1}^{2}
a_{1}^{2} c^{- 2}(1 - e_{1}^{2})^{- 1}) \nonumber \\
+ (1.0118a_{3}a_{1}^{- 1}\sin (\phi_{3;0} + 2.3544 + 99\pi
\omega_{3}^{2} a_{3}^{2} c^{- 2}(1 - e_{3}^{2})^{- 1})) \nonumber \\
+ 0.01485(1 - e_{1})^{2}\sin \phi_{1;0} \sin (\phi_{1;0} +
415\pi \omega_{1}^{2} a_{1}^{2} c^{- 2}(1 - e_{1}^{2})^{- 1}))) \nonumber \\
\times (((1 - e_{1})\cos \phi_{1;0} - 1.0157a_{3}a_{1}^{- 1}
\cos (\phi_{3;0} + 2.7521))^{2} \nonumber \\
+ (0.99255(1 - e_{1})\sin \phi_{1;0} + 1.0157a_{3}a_{1}^{- 1}\sin
(\phi_{3;0} + 2.7521))^{2} \nonumber
\\ + 0.01485(1 - e_{1})^{2}\sin^{2} \phi_{1;0})^{- 1/2}
\nonumber \\ \times (((1 - e_{1})\cos (\phi_{1;0} + 415\pi \omega_{1}^{2}
a_{1}^{2} c^{- 2}(1 - e_{1}^{2})^{- 1}) \nonumber
\\ - 1.0118a_{3}a_{1}^{- 1}\cos ( \phi_{3;0} + 2.3544 + 99
\pi \omega_{3}^{2} a_{3}^{2} c^{- 2}(1 - e_{3}^{2})^{- 1}))^{2}
\nonumber \\ + ( 0.99255(1 - e_{1})\sin (\phi_{1;0} + 415\pi
\omega_{1}^{2} a_{1}^{2} c^{- 2}(1 - e_{1}^{2})^{- 1}) \nonumber
\\ + 1.0118a_{3}a_{1}^{- 1}\sin (\phi_{3;0} + 2.3544 + 99\pi
+ \omega_{3}^{2} a_{3}^{2} c^{- 2}(1 - e_{3}^{2})^{- 1}))^{2} \nonumber \\
+ 0.01485(1 - e_{1})^{2}\sin^{2} ( \phi_{1;0} + 415\pi \omega_{1}^{2}
a_{1}^{2} c^{- 2}(1 - e_{1}^{2})^{- 1}))^{- 1/2}.
\end{eqnarray}
The perihelion angles $\phi_{1;0}$, $\phi_{3;0}$ are needed. Let the
perihelion angles $\phi_{1;0}$, $\phi_{3;0}$ in the equation (\ref{2.45}) be
equal to zero. Then $\alpha (0,415) = 17^{o}.889$. According to (Ref. 5, Chap.
40, Sec. 40.5, Appendix 40.3), the advance of Mercury's perihelion, observed
by the astronomers from the Earth, is $1^{o}.55548 \pm 0^{o}.00011$ per century.

\section{III. GENERAL RELATIVITY PLANET ORBITS}
\setcounter{equation}{0}

The equations (\ref{1.12}) are the Kepler problem equations for the proper time $\tau$.
Due to the equations (\ref{1.12}) the angular momentum
\begin{equation}
\label{3.10} M_{l}({\bf x}) = \sum_{i,j = 1}^{3} \epsilon_{ijl}
\left( x^{i}\frac{dx^{j}}{d\tau} -
x^{j}\frac{dx^{i}}{d\tau} \right),\, \, l = 1,2,3,
\end{equation}
and the energy
\begin{equation}
\label{3.11} E ({\bf x}) = \frac{1}{2} \Biggl| \frac{d{\bf x}}{d\tau} \Biggr|^{2} -
\frac{m_{10}G}{|{\bf x}|}
\end{equation}
are independent of the proper time $\tau$. The antisymmetric in
all indices tensor $\epsilon_{ij\, l}$ has the normalization
$\epsilon_{123} = 1$. Let the third axis coincide with
the constant vector (\ref{3.10}). The vector ${\bf x} (\tau)$ is
orthogonal to the constant vector (\ref{3.10}): $x^{3} (\tau) =  0$.
Inserting the vector
\begin{equation}
\label{3.12} x^{1} (\tau) = r(\tau) \cos \phi (\tau),\, \,
x^{2} (\tau) = r(\tau) \sin \phi (\tau), \, \, x^{3} (\tau) =  0
\end{equation}
into the relations (\ref{3.10}), (\ref{3.11}) we get for the angular momentum
\begin{equation}
\label{3.13} |{\bf M} ({\bf x})| = r^{2}(\tau) \frac{d\phi}{d\tau}
\end{equation}
and for the energy
\begin{equation}
\label{3.14} E ({\bf x}) = \frac{1}{2} \left( \left( \frac{dr}{d\tau} \right)^{2} +
r^{2}(\tau) \left( \frac{d\phi}{d\tau} \right)^{2} \right) - \frac{m_{10}G}{r(\tau)}.
\end{equation}
It follows from the equations (\ref{3.13}), (\ref{3.14}) that
\begin{equation}
\label{3.15} \left( \frac{dr}{d\tau} \right)^{2} = 2E ({\bf x}) + 2\frac{m_{10}G}{r}
- \frac{|{\bf M} ({\bf x})|^{2}}{r^{2}}.
\end{equation}
Dividing the equation (\ref{3.15}) by the square of the equation (\ref{3.13}) we get
\begin{equation}
\label{3.16}  \left(  \frac{d}{d\phi} \, \, \frac{1}{r} \right)^{2} =
\frac{2E ({\bf x})}{|{\bf M} ({\bf x})|^{2}}  + \frac{2m_{10}G}{|{\bf M} ({\bf x})|^{2} r}
- \frac{1}{r^{2}}.
\end{equation}
The function
\begin{equation}
\label{3.17}
a (1 - e^{2})r^{- 1} =
1 + e \cos \left( \phi - \phi_{0} \right),
\end{equation}
$$
a (1 - e^{2}) = m_{10}^{- 1}G^{- 1} |{\bf M} ({\bf x}))|^{2},
$$
$$
e^{2} = 1 +  2m_{10}^{- 2}G^{- 2}E ({\bf x})|{\bf M} ({\bf x}))|^{2}
$$
is the solution of the equation (\ref{3.16}). $\phi_{0}$  is the perihelion angle.
We assume that the constants (\ref{3.10}), (\ref{3.11})  satisfy the inequality
\begin{equation}
\label{3.20} 0 < - 2E ({\bf x}))|{\bf M} ({\bf x}))|^{2} \leq m_{10}^{2}G^{2}.
\end{equation}
The third equality (\ref{3.17}) and the inequality (\ref{3.20}) imply the inequality
$0 \leq e^{2} < 1$ for the orbit eccentricity. The relations (\ref{3.13}), (\ref{3.17})
imply the angle $\phi (\tau)$ dependence on $\tau$
\begin{equation}
\label{3.1} \tau = \tau_{0} + \int_{\phi_{0}}^{\phi} d\psi
\frac{a^{2}(1 - e^{2})^{2}|{\bf M} ({\bf x}))|^{- 1}}{\left( 1 +
e \cos \left( \psi - \phi_{0} \right) \right)^{2}}.
\end{equation}
Let us consider the metric (\ref{1.7}). The world line $x^{\mu} (t)$ is called
geodesic, if it satisfies the geodesic equations (Ref. 5, Chap. 13, Sec. 13.4,
equations (13.36)) with the proper time $\tau$ (the second relation (\ref{1.12}))
\begin{equation}
\label{3.2} \sum_{\nu \, =\, 0}^{3} g_{\sigma \nu} (x)
\frac{d^{2}x^{\nu}}{d\tau^{2}}  = - \frac{1}{2} \sum_{\mu, \nu \, = \, 0}^{3}
\left( \frac{\partial g_{\sigma \nu} (x)}{\partial x^{\mu}}
+ \frac{\partial g_{\sigma \mu} (x)}{\partial x^{\nu}}
- \frac{\partial g_{\mu \nu} (x)}{\partial x^{\sigma}} \right)
\frac{dx^{\mu}}{d\tau} \frac{dx^{\nu}}{d\tau},
\end{equation}
$\sigma = 0,1,2,3$. The general relativity planet orbit is the geodesic world line
for the metric (\ref{1.7}). The second relation (\ref{1.12}) implies the identity
(Ref. 5, Chap. 13, Sec. 13.4, relation (13.37))
\begin{equation}
\label{3.3} \sum_{\mu, \nu \, = \, 0}^{3} g_{\mu \nu} (x)
\frac{dx^{\mu}}{d\tau} \frac{dx^{\nu}}{d\tau} = c^{2}
\end{equation}
similar to the first identity (\ref{1.19}). The differentiation of the identity
(\ref{3.3}) yields the identity
$$
\sum_{\sigma, \nu \, =\, 0}^{3}
g_{\sigma \nu} (x) \frac{dx^{\sigma}}{d\tau}
\frac{d^{2}x^{\nu}}{d\tau^{2}}  =
$$
\begin{equation}
\label{3.4} - \frac{1}{2} \sum_{\sigma, \mu, \nu \, = \, 0}^{3}
\left( \frac{\partial g_{\sigma \nu}
(x)}{\partial x^{\mu}}
+ \frac{\partial g_{\sigma \mu} (x)}{\partial x^{\nu}}
- \frac{\partial g_{\mu \nu} (x)}{\partial x^{\sigma}} \right)
\frac{dx^{\sigma}}{d\tau} \frac{dx^{\mu}}{d\tau} \frac{dx^{\nu}}{d\tau}.
\end{equation}
Therefore the geodesic equation (\ref{3.2}), $\sigma = 0$ is a linear combination of the
geodesic equations (\ref{3.2}), $\sigma = 1,2,3$. We consider these equations for
the world line $x^{\mu} (t)$, $x^{0} (t) = ct$,
$$
\left( 1 + 2 \frac{m_{10}G}{rc^{2}} \right) \left( 1 - 2\frac{m_{10}G}{rc^{2}} +
2\frac{m_{10}^{2}G^{2}}{r^{2}c^{4}}
- \left( 1 + 2\frac{m_{10}G}{rc^{2}}\right) \frac{1}{c^{2}}
\Biggl| \frac{d{\bf x}}{dt} \Biggr|^{2} \right) \frac{d^{2}x^{i}}{d\tau^{2}}  =
$$
\begin{equation}
\label{3.8}
- \frac{m_{10}G}{r^{2}} \left( \frac{x^{i}}{r} \left( 1 - 2\frac{m_{10}G}{rc^{2}} +
\frac{1}{c^{2}} \Biggl| \frac{d{\bf x}}{dt}\Biggr|^{2} \right) - \frac{4}{c^{2}}
\frac{dx^{i}}{dt} \frac{dr}{dt} \right), \, \, i = 1,2,3.
\end{equation}
The geodesic equations (\ref{3.8}) without the terms
\begin{equation}
\label{3.21}
\pm \, \, 2\frac{m_{10}G}{rc^{2}}, \, \, 2\frac{m_{10}^{2}G^{2}}{r^{2}c^{4}}, \, \,
\pm \, \, \frac{1}{c^{2}} \Biggl| \frac{d{\bf x}}{dt}\Biggr|^{2}, \, \,
- \, \frac{4}{c^{2}} \frac{dx^{i}}{dt} \frac{dr}{dt}, \, \, i = 1,2,3,
\end{equation}
coincide with the Kepler problem equations (\ref{1.12}). The equations (\ref{1.12}) are
solved exactly. Let us insert the solution (\ref{3.12}), (\ref{3.17}) of the equations
(\ref{1.12}) into the geodesic equations (\ref{3.8}) and estimate the terms (\ref{3.21})
\begin{equation}
\label{4.13}
\frac{m_{10}G}{r_{k}c^{2}} =
\omega_{k}^{2} a_{k}^{2}c^{ - 2} \,
\frac{1 + e_{k} \cos \left( \phi_{k} - \phi_{k;\, 0} \right)}{1 - e_{k}^{2}} \, \ll 1,
\end{equation}
\begin{equation}
\label{3.23} \frac{m_{10}^{2}G^{2}}{r_{k}^{2}c^{4}} =
\omega_{k}^{4} a_{k}^{4}c^{ - 4} \,
\frac{(1 + e_{k} \cos \left( \phi_{k} - \phi_{k;\, 0} \right) )^{2}}{(1 - e_{k}^{2})^{2}} \, \ll 1,
\end{equation}
\begin{equation}
\label{3.22}
\frac{1}{c^{2}} \Biggl| \frac{d{\bf x}_{k}}{dt}\Biggr|^{2} \approx
\frac{a_{k}^{2}\omega_{k}^{2}c^{- 2}(1 - e_{k}^{2})^{2}}{\left( 1 +
e_{k} \cos \left( \phi_{k} -
\phi_{k;\, 0} \right) \right)^{2}}
\left( \frac{ e_{k}^{2}\sin^{2} \left( \phi_{k} -
\phi_{k;\, 0} \right)}{\left( 1 + e_{k} \cos \left( \phi_{k} -
\phi_{k;\, 0} \right) \right)^{2}} + 1 \right) \ll 1,
\end{equation}
$$
\frac{4}{c^{2}} \left( \sum_{i\, =\, 1}^{3}
\left( \frac{dx_{k}^{i}}{dt} \right)^{2} \right)^{1/2}
\Biggl| \frac{dr_{k}}{dt} \Biggr|
\approx \frac{ 4\omega_{k}^{2}a_{k}^{2}c^{- 2}e_{k}(1 - e_{k}^{2})^{2}
|\sin \left( \phi_{k} - \phi_{k;\, 0} \right) |}{\left( 1 +
e_{k} \cos \left( \phi_{k} - \phi_{k;\, 0} \right)\right)^{3}} \, \times
$$
\begin{equation}
\label{3.24}
\left( \frac{ e_{k}^{2}\sin^{2} \left( \phi_{k} - \phi_{k;\, 0} \right)}{\left( 1 +
e_{k} \cos \left( \phi_{k} - \phi_{k;\, 0} \right) \right)^{2} } + 1 \right)^{1/2}
\, \, \ll \left( \sum_{i\, =\, 1}^{3}
\left( \frac{x_{k}^{i}}{r_{k}} \right)^{2} \right)^{1/2} = 1,
\end{equation}
$k = 1,...,9$. We assumed that $m_{10}Gc^{- 2} \approx \omega_{k}^{2} a_{k}^{3}c^{- 2}$
(the third Kepler law (\ref{1.10})) and $|d\phi_{k} / dt| \, \approx \omega_{k} = 2\pi
T_{k}^{- 1}$. According to (Ref. 5, Chap. 25, Sec. 25.1, Appendix 25.1) the values
$\omega_{k}^{2} a_{k}^{3}c^{- 2} = 1477m$ for $k = 1,2,3,4,6$
(Mercury, Venus, the Earth, Mars and Saturn), the values
$\omega_{l}^{2} a_{l}^{3}c^{- 2} = 1478m$ for $l = 5,8$ (Jupiter and Neptune),
the value $\omega_{7}^{2} a_{7}^{3}c^{- 2} = 1476m$ for Uranus, the value
$\omega_{9}^{2} a_{9}^{3}c^{- 2} = 1469m$ for Pluto. The major semi-axes:
$a_{1} = 0.5791\cdot 10^{11}m$, $a_{2} = 1.0821\cdot 10^{11}m$,
$a_{3} = 1.4960\cdot 10^{11}m$, $a_{4} = 2.2794\cdot 10^{11}m$,
$a_{5} = 7.783\cdot 10^{11}m$, $a_{6} = 14.27\cdot 10^{11}m$,
$a_{7} = 28.69\cdot 10^{11}m$, $a_{8} = 44.98\cdot 10^{11}m$,
$a_{9} = 59.00\cdot 10^{11}m$. For Mercury
$\omega_{1}^{2}a_{1}^{2}c^{- 2} \approx 2.6 \cdot 10^{- 8}$, for Venus
$\omega_{2}^{2}a_{2}^{2}c^{- 2} \approx 1.4 \cdot 10^{- 8}$, for the Earth
$\omega_{3}^{2}a_{3}^{2}c^{- 2} \approx 9.9 \cdot 10^{- 9}$, for Mars
$\omega_{4}^{2}a_{4}^{2}c^{- 2} \approx 6.5 \cdot 10^{- 9}$, for Jupiter
$\omega_{5}^{2}a_{5}^{2}c^{- 2} \approx 1.9 \cdot 10^{- 9}$, for Saturn
$\omega_{6}^{2}a_{6}^{2}c^{- 2} \approx 1.04 \cdot 10^{- 9}$, for Uranus
$\omega_{7}^{2}a_{7}^{2}c^{- 2} \approx 5.1 \cdot 10^{- 10}$, for Neptune
$\omega_{8}^{2}a_{8}^{2}c^{- 2} \approx 3.3 \cdot 10^{- 10}$, for Pluto
$\omega_{9}^{2}a_{9}^{2}c^{- 2} \approx 2.5 \cdot 10^{- 10}$. The orbit
eccentricities: $e_{1} = 0.21$, $e_{2} = 0.007$, $e_{3} = 0.017$,
$e_{4} = 0.093$, $e_{5} = 0.048$, $e_{6} = 0.056$, $e_{7} = 0.047$,
$e_{8} = 0.009$, $e_{9} = 0.249$.

For the world line $x^{\mu} (t)$, $x^{0} (t) = ct$, the second relation (\ref{1.12}) is
\begin{equation}
\label{4.14}
\frac{dt}{d\tau} = \left( 1 - 2\frac{m_{10}G}{rc^{2}} +
2\frac{m_{10}^{2}G^{2}}{r^{2}c^{4}}
- \left( 1 + 2\frac{m_{10}G}{rc^{2}} \right) \frac{1}{c^{2}}
\Biggl| \frac{d{\bf x}}{dt} \Biggr|^{2} \right)^{- 1/2}.
\end{equation}
The vector ${\bf x} (\tau)$ is the solution of the equations (\ref{1.12}). It depends on
the proper time $\tau$. Then the equality (\ref{4.14}) implies
\begin{equation}
\label{4.15} \left( 1 - 2\frac{m_{10}G}{rc^{2}} +
2\frac{m_{10}^{2}G^{2}}{r^{2}c^{4}} \right) \left( \frac{dt}{d\tau} \right)^{2} =
1 + \left( 1 + 2\frac{m_{10}G}{rc^{2}} \right) \frac{1}{c^{2}}
\Biggl| \frac{dt}{d\tau} \frac{ d{\bf x}}{dt} \Biggr|^{2}.
\end{equation}
The equation (\ref{4.15}) has the solution
$$
t(\tau) = t(0)\, \, + \, \, \int_{0}^{\tau} d \tau^{\prime} \left( 1 + \left( 1 +
2\frac{m_{10}G}{r(\tau^{\prime})c^{2}} \right) \frac{1}{c^{2}}
\Biggl| \frac{dt}{d\tau^{\prime}} \frac{ d{\bf x}}{dt} \Biggr|^{2} \right)^{1/2} \times
$$
\begin{equation}
\label{4.16}
\left( 1 - 2\frac{m_{10}G}{r(\tau^{\prime})c^{2}} +
2\frac{m_{10}^{2}G^{2}}{r^{2}(\tau^{\prime})c^{4}} \right)^{- 1/2},
\end{equation}
$$
g_{00} (x) \, \, = 1 - 2\frac{m_{10}G}{rc^{2}} +
2\frac{m_{10}^{2}G^{2}}{r^{2}c^{4}} \, \, =
\left( 1 - \frac{m_{10}G}{rc^{2}} \right)^{2}
+ \frac{m_{10}^{2}G^{2}}{r^{2}c^{4}} \, \, \geq \, \, \frac{1}{2}.
$$
The relations (\ref{3.12}), (\ref{3.17}), (\ref{3.1}), (\ref{4.16}) are the
complete description of planet orbit. It is the orbit (\ref{1.11}), (\ref{1.5}) with
the precession coefficient $\gamma = 1$.

By making use of the relation (\ref{4.14}) and of the estimates similar to (\ref{4.13})
- (\ref{3.22}) it is possible to prove for the solution (\ref{1.11}) - (\ref{1.6}) of
the equation (\ref{1.13})
\begin{equation}
\label{3.18}
\frac{dt}{d\tau} \frac{d}{dt} \left( \frac{dt}{d\tau} \frac{d{\bf x}}{dt}\right)
\approx \frac{d}{dt} \left( \left( 1 - \frac{1}{c^{2}}
\Biggl| \frac{d{\bf x}}{dt} \Biggr|^{2} \right)^{- 1/2} \frac{d{\bf x}}{dt}\right)
= - \frac{m_{10}G}{|{\bf x}|^{3}} \, {\bf x}.
\end{equation}
Due to the relation (\ref{3.18}) the solution (\ref{1.11}) - (\ref{1.6}) of
the equation (\ref{1.13}) is an approximate solution of the equation (\ref{1.12}) for the
world line $x^{\mu} (t)$, $x^{0} (t) = ct$.

Let us compare the function (\ref{1.5}), (\ref{1.6}) with the function (\ref{1.5}),
(\ref{1.10}), (\ref{1.8}) from (Ref. 5, Chap. 40, Sec. 40.5)
$$
1 + \, e \cos \left( \left( 1 - \frac{3\omega^{2} a^{2}}{(1 - e^{2})c^{2}} \right)
(\phi (t) - \phi_{0} )\right) =  1 + \, e \cos \left( \left( 1 - \frac{\omega^{2} a^{2}}{2(1 -
e^{2})c^{2}} \right) (\phi (t) - \phi_{0} )\right)
$$
\begin{equation}
\label{3.19}  +  \, \frac{e\omega^{2} a^{2}(\phi (t) - \phi_{0} )}{(1 - e^{2})c^{2}}
\int_{1/2}^{3} ds \, \sin \, \left( \left( 1 - \frac{s\omega^{2} a^{2}}{(1 - e^{2})c^{2}}
\right) (\phi (t) - \phi_{0} )\right) \, .
\end{equation}
The value $\omega^{2} a^{2}c^{- 2}$ is negligeable for any planet.

\vskip 0.5cm

\noindent {\bf ACKNOWLEDGMENTS}

\vskip 0.5cm

This work was supported in part by the Program for Supporting Leading Scientific Schools
(Grant No. 4612.2012.1) and by the RAS Program "Fundamental Problems of Nonlinear
Mechanics."

\vskip 0.5cm

\noindent ${}^{1}$Poincar\'e, H., "Sur la structure d'\'electron," Rendiconti Circolo Mat.
Palermo. {\bf 21}, 129 - 176 (1906); {\it Oeuvres de Henri Poincar\'e}, t. IX (Gauthier - Villars,
Paris, 1956), p. 494 - 550.

\noindent ${}^{2}$Lorentz, H. A., "Electromagnetic phenomena in a system moving
with any velocity smaller than that of light," Proc. Roy. Acad. Amsterdam. {\bf 6},
809 - 831 (1904).

\noindent ${}^{3}$Zinoviev, Yu.M., "Gravity and Lorentz Force," Theor. Math. Phys., {\bf
131}, 729 - 746 (2002).

\noindent ${}^{4}$Sommerfeld, A.: {\it Elektrodynamik} (Akademische Verlagsgesellschaft
Geest $\&$ Portig K.-G., Leipzig, 1949).

\noindent ${}^{5}$Misner, C.W., Thorne, K.S. and Wheeler, J.A., {\it Gravitation} (Freeman, San
Francisco, 1973).

\noindent ${}^{6}$Clemence, G.M., "The Relativity Effect in Planetary Motions," Rev. Mod. Phys.
{\bf 19}, 361 - 364 (1947).

\noindent ${}^{7}$Boguslavskii, S.A., {\it Selected Works on Physics} (in Russian) (State
Publishing House for Literature on Physics and Mathematics, Moscow, 1961).

\noindent ${}^{8}$Einstein, A., "Erkl\"arung der Perihelbewegung der Merkur aus der
allgemeinen

\noindent Relativit\"atstheorie,"  Sitzungsber. preuss. Akad. Wiss. {\bf 47}, 831 - 839
(1915).

\noindent ${}^{9}$Eddington, A.S., {\it The Mathematical Theory of Relativity} \, (Cambridge
University Press, Cambridge, UK, 1924).

\noindent ${}^{10}$Zinoviev, Yu.M., "Causal electromagnetic interaction equations," J. Math. Phys.
{\bf 52}, 022302 (2011).

\noindent ${}^{11}$Vladimirov, V.S., {\it Methods of Theory of Many Complex Variables} (MIT
Press, Cambridge, MA, 1966).

\noindent ${}^{12}$Einstein, A., "Einheitliche Feldtheorie von Gravitation und Elektrizit\"at,"
Sitzungsber. preuss. Akad. Wiss., phys.-math. Kl. 414 - 419 (1925).

\end{document}